\documentclass[acus]{JAC2003}

\usepackage{graphicx, epsfig, subfigure,bm,amssymb}
\usepackage{amssymb}
\usepackage{amsmath}
\usepackage{mathptmx,times}

\def\Im{\mathop{{\cal I}\! m}}
\def\Re{\mathop{{\cal R}\! e}}

\def\case#1/#2{\textstyle\frac{#1}{#2}}

\def\beq{\begin{equation}}
\def\eeq{\end{equation}}
\def\bdis{\begin{displaymath}}
\def\edis{\end{displaymath}}

\begin{document}
\title{IMPEDANCES OF TEVATRON \vspace{-0.12in}SEPARATORS}
\author{
K.Y.~Ng,\thanks{Email: ng@fnal.gov.
Work supported by the U.S. Department of Energy under contract 
No.~DE-AC02-76CH03000.} 
FNAL, Batavia, IL 60510, USA}

\maketitle

\begin{abstract}\vskip-0.05in
The impedances of the Tevatron separators are revisited and are found
to be negligibly small in the few hundred MHz region, 
except for resonances at 22.5~MHz.  The latter
are contributions from the power cables which may drive head-tail instabilities
if the bunch is long enough.\\[-0.24in]
\end{abstract}

\section{I.~~INTRODUCTION\vspace{-0.05in}}

Large chromaticities ($\sim12$ units) were required to control the vertical
transverse head-tail instabilities observed recently in the Tevatron.
Application of the head-tail growth expression~\cite{alex} reveals
that the necessary transverse impedance to drive such instabilities
has to be at least twice the amount estimated in the 
Run~II Handbook.~\cite{runii}
This underestimation becomes thrice when it was suggested~\cite{burov-lebedev}
that the transverse impedance of the Lambertson magnets have been
overestimated by ten fold.\footnote{
The Run~II Handbook estimate has been rather rough, but
reasonable.  
The C0 Lambertson magnet removed recently shows very large transverse
impedance.}
It was further suggested that most of the transverse impedance actually
comes from the static separators: the vertical transverse impedance should be
5.33~M$\Omega$/m assuming 27 separators
while the Run~II Handbook estimate has been
only 0.082~M$\Omega$/m assuming 11 separators.\footnote{
0.82~M$\Omega$/m quoted in the Run~II Handbook is a misprint.}
  This 26-time
difference for each separator
prompts us to review
the impedances in detail by numerical computation, theoretical reasoning,
and experimental measurement.
The conclusion points to the fact that the separators actually contribute
negligibly when compared with other
discontinuities in the Tevatron vacuum chamber, 
except for the rather large resonances at 22.5~MHz due to the power
\vspace{-0.08in}cables.

\section{II.~~NUMERICAL COMPUTATIONS\vspace{-0.05in}}

\begin{figure}[h]
\centering{\epsfig{figure=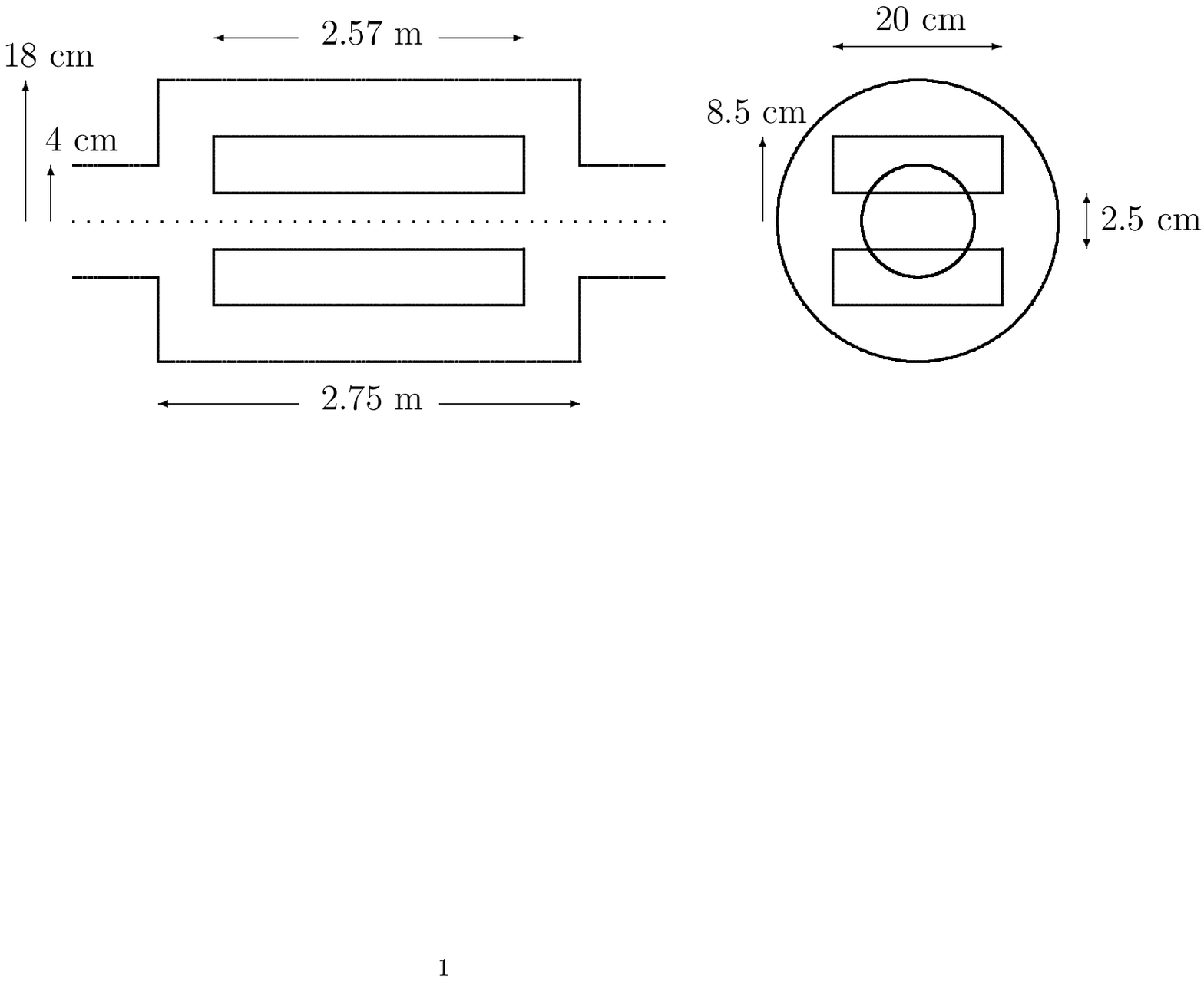,width=3.2in,bbllx=130bp,bblly=305bp,bburx=610bp,bbury=500bp,clip=}}
\vskip-0.18in
\caption{\small The simplified separator model used in MAFIA computation of
longitudinal and transverse wake potentials.}\label{separator}
\vskip-.05in
\centering{\epsfig{figure=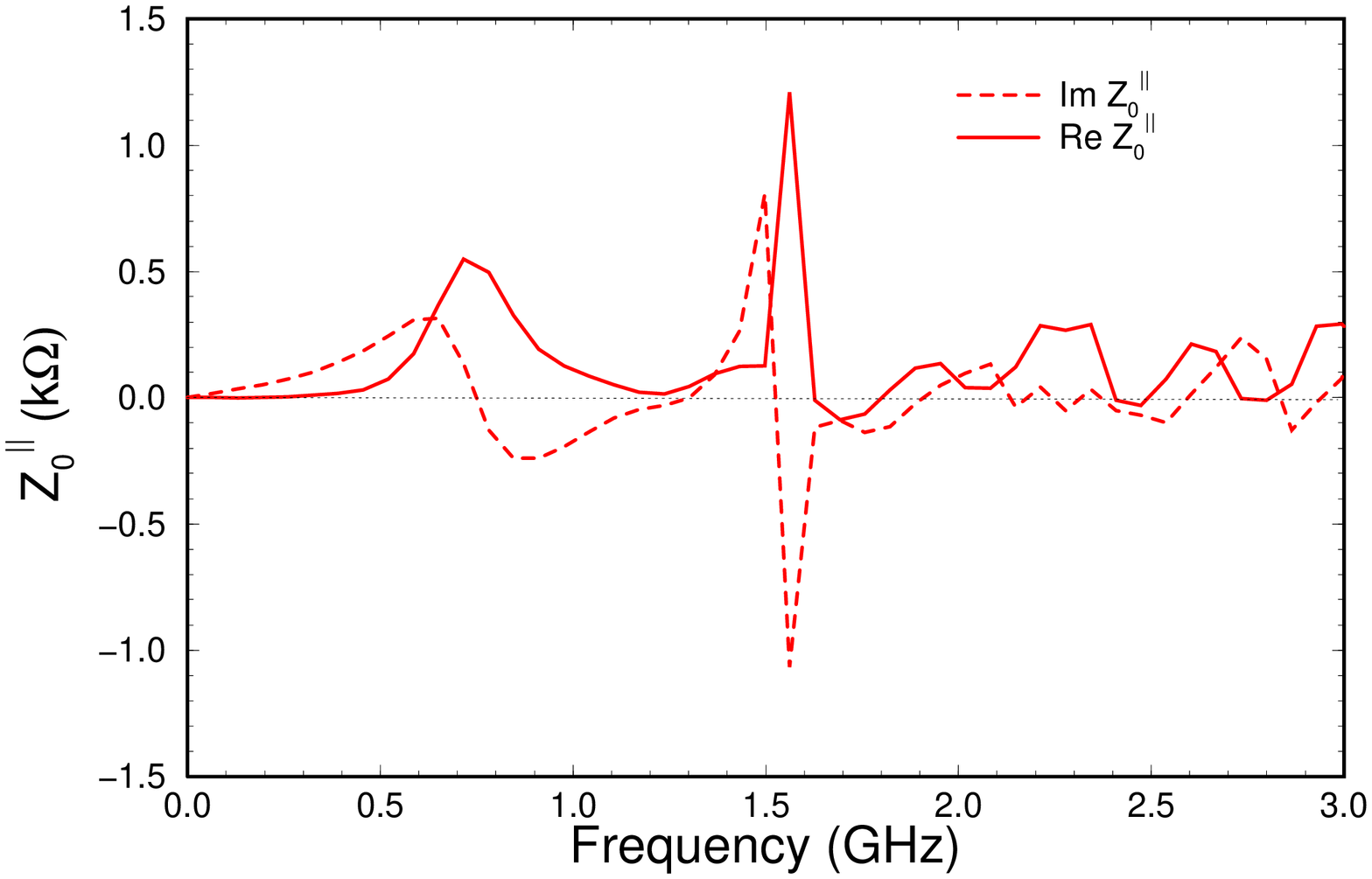,height=3.1in,width=1.85in,
angle=-90,
bbllx=180bp,bblly=45bp,bburx=565bp,bbury=635bp,clip=}}
\vskip-.20in
\centering{\epsfig{figure=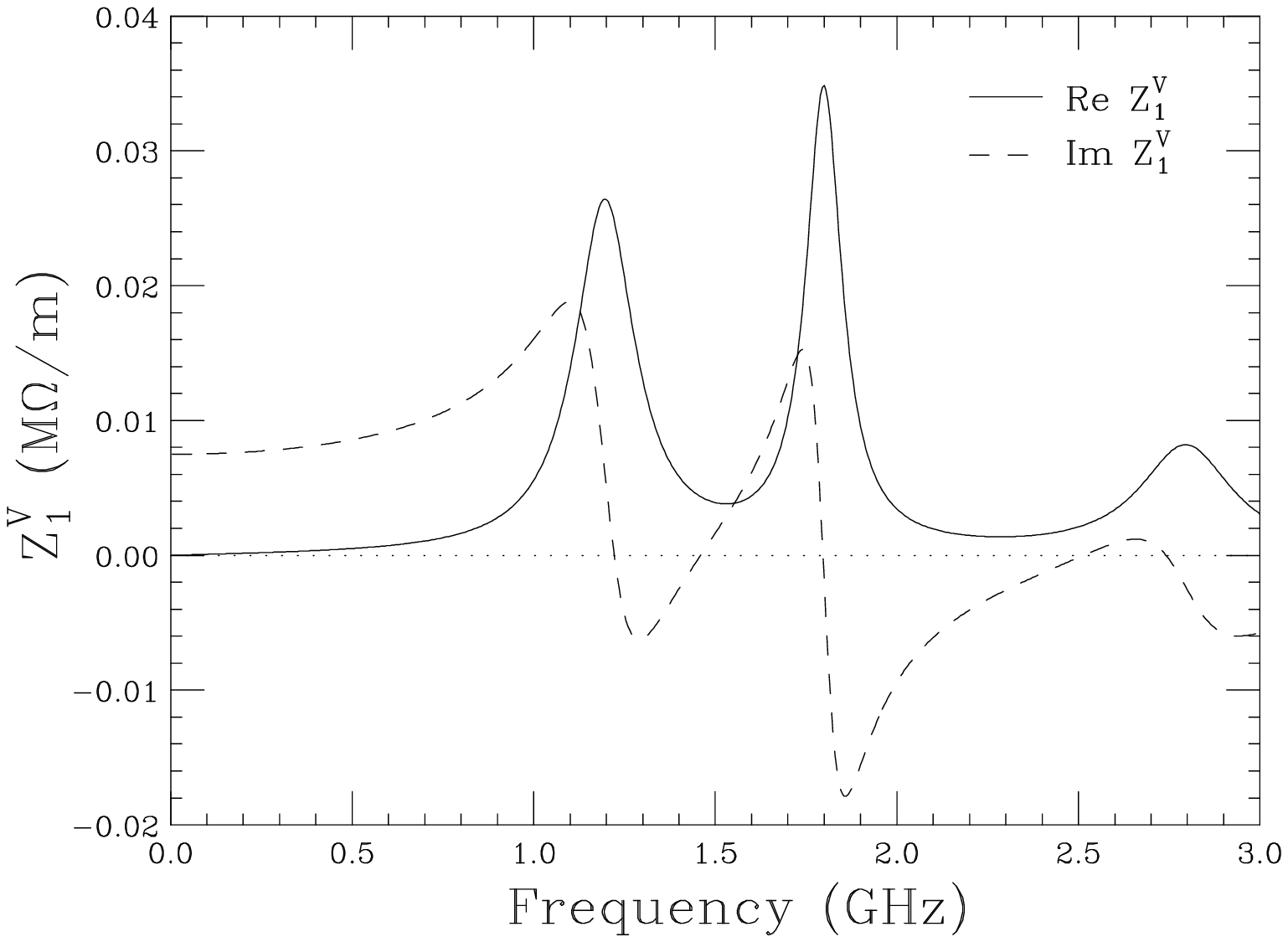,height=2.0in,width=3.1in,bbllx=75bp,bblly=298bp,bburx=513bp,bbury=657bp,clip=}}
\vskip-.15in
\caption{\small The real and imaginary parts of $Z_0^\parallel$ (top)
and 
$Z_1^V$ (bottom) of one separator as computed by MAFIA.}\label{mafia}
\vskip-.24in
\end{figure}

We model a separator without the power cables 
as two plates 6~cm thick,
20~cm wide, 2.57~m long, separated by 5~cm inside a circular 
chamber of length 2.75~m and
radius 18~cm (Fig.~\ref{separator}).  
The beam pipe is circular in
cross section with radius 4~cm.
The 3-D code MAFIA in the time domain~\cite{mafia} 
has been used to obtain the longitudinal
and transverse impedances shown 
in Fig.~\ref{mafia}.
We find that, at low frequencies, the longitudinal impedance per harmonic and
the vertical transverse impedance are, respectively,
$Z_0^\parallel/n\sim 0.019j~\Omega$ and
 $Z_1^V\sim 0.0075j~{\rm M}\Omega$/m, which agree
with the estimates given in the Run~II Handbook. 

%\section{ANALYSIS}

\begin{figure}[h]
\centering{\epsfig{figure=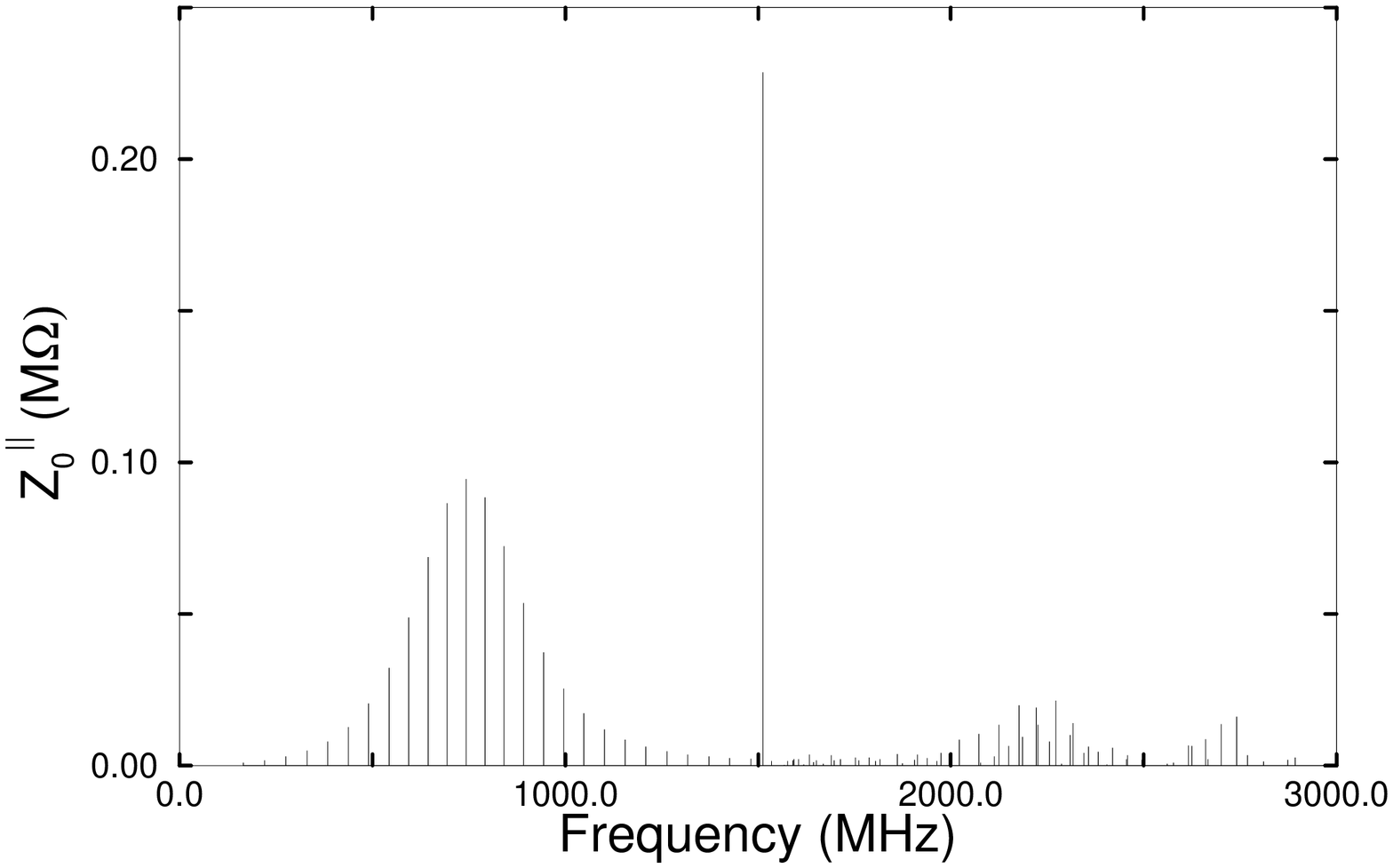,
height=3.1in,width=1.83in,angle=-90,
bbllx=190bp,bblly=35bp,bburx=560bp,bbury=648bp,clip=}}
\vskip0.05in
\centering{\epsfig{figure=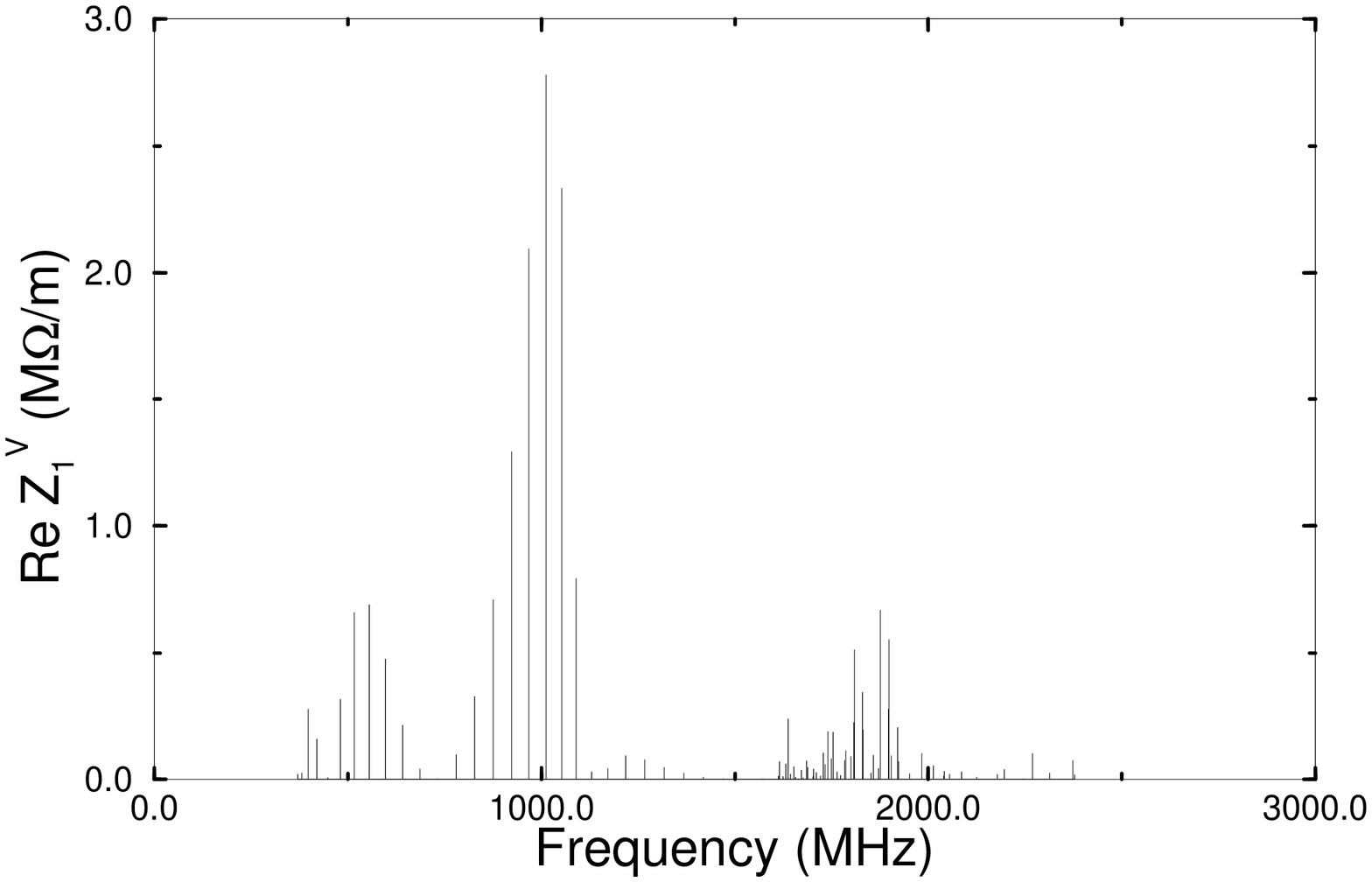,
height=3.1in,width=1.83in,angle=-90,
bbllx=190bp,bblly=35bp,bburx=565bp,bbury=648bp,clip=}}
\vskip-0.1in
\caption{\small $\Re Z_0^\parallel$ (top) and $\Re Z_1^V$ (bottom)
as obtained from 2-D URMEL computation.}\label{urmel}
\vskip-.2in
\end{figure}

In many cases, a 2-D approximation, assuming cylindrical symmetry
of the 3-D problem, should give us a good insight as to
the physics of the problem.  
The advantage is obvious; we can use more mesh points to better represent
the geometry.
The first 50 resonant modes computed by the 2-D URMEL code~\cite{mafia} are
shown in Fig.~\ref{urmel}~(top).  They are well below the cutoff frequency of
4.59~GHz and therefore appear as narrow resonances.
The separator can be viewed as two pill-box cavities joined by a
coaxial waveguide.  The coaxial waveguide resonates when its length equals
to an integral number of half wavelengths.  Thus, the lowest mode has a
frequency of $c/(2\ell)=54.5$~MHz and successive modes are separated also
by 54.5~MHz, where $c$ is the velocity of light and $\ell=2.75$~m is the
length of separator.  
To excite these standing TEM modes
in the coaxial waveguide, electromagnetic fields must penetrate into the
separator, and penetration is only efficient when the frequency is
near the resonant frequencies of the cavities at each end of the
separator.  
These pill-box-like cavities have a radial depth of $d=18$~cm
with the first resonance at $637$~MHz,
and we do see coaxial-guide modes excited very much stronger
near this frequency.  The next pill-box-cavity mode is at 1463~MHz
with a radial node at $7.84$~cm which is very near the outer
radius of 8.5~cm of the cylindrical plates.  Thus a rather pure cavity mode
is excited with very little contamination from the coaxial guide.  This
explains why we see a very strong excitation of this mode without many
coaxial-guide modes at nearby frequencies.  The third pill-box-cavity
mode at 2294~MHz can also be
seen in the figure with coaxial-guide modes at surrounding frequencies.
Because excitation decreases with frequency, the shunt impedances
are much smaller.

Due to the finite mesh size and rms bunch length used in the computation,
MAFIA broadens all these sharp resonances.  
If all quality factors are broadened to $Q=15$,  
the results look very similar to
those in Fig.~\ref{mafia}, implying that our interpretation of the longitudinal
impedance of the separator is correct.

Similar analysis applies to the transverse dipole modes.  
The lowest 50 resonances computed by URMEL are shown in  
Fig.~\ref{urmel}~(bottom).
The first two
transverse resonances in the pill-box cavities are
1016, 1860~MHz.
We do see coaxial-guide modes enhanced near these frequencies.
There is a special mode when one wavelength of the magnetic field
wraps around the ``cylindrical plate'' between the plate and the
encasing outer shield.  The radius is from $r=8.5$ to 18~cm.
The wavelength will be $\lambda=2\pi r$ and the frequency will be
between 265 and 562~MHz.  This explains the
cluster of low-frequency coaxial-guide modes
in the URMEL results.  There is no cylindrical symmetry in the
actual separator and this low-frequency cluster is
therefore not present in the MAFIA results.
Again if we broaden the sharp resonances until the quality factor
reaches 15, the real and imaginary parts of the transverse
impedance will look
similar to the MAFIA results of\vspace{-0.1in} Fig.~\ref{mafia}.

\section{III.~~COMPARISON WITH BPM\vspace{-0.05in}}

Although the Tevatron stripline beam-position monitor (BPM) is similar in
structure to the separator,
however, its impedance is completely different.
Here, the striplines play the role of
the separator plates.
The main difference is that each end of the
striplines is terminated with a resistor of $Z_c=50~\Omega$, which is equal to
the characteristic impedance of the transmission line formed from the 
stripline and the enclosing outer shield.
As a pulse of protons crosses the upstream gap, it creates
on the beam-side of the stripline negative image charges, which
move forward with the beam pulse.  Since the stripline is neutral, positive
charges will be created at the underside of the stripline.  These
positive charges, seeing a termination $Z_c$ in parallel with a transmission
line of characteristic impedance $Z_c$, split into two equal halves:
one half flows down the termination while the other half travels along
the transmission line and flows down the termination $Z_c$ at the other end
of the stripline.  When the beam pulse crosses the downstream gap
of the BPM, the negative image charges on the beam-side of the stripline
wrap into the underside of the stripline;
one half flows down the downstream
termination 
while the other half
flows backward along the transmission line and eventually
down the upstream termination.  Assuming  
the transmission line velocity 
to be the same as the beam velocity, the current in
the downstream termination vanishes between one half of the stripline
underside positive charges and one half of the wrap-around negative
image charges.  At the upstream termination, we see first a
positive signal followed by a negative signal delayed by twice the transit
time of traveling along the stripline.
Thus the potential across the upstream gap 
\vspace{-0.1in}is
\beq
V(t)=\frac12Z_c\left[I_0(t)-I_0(t-2\ell/c)\right]~,
\label{bpm}
\vspace{-0.1in}
\eeq
where $\ell$ is the length of the stripline and $I_0(t)$ is the beam
current. 
The factor $\frac12$ occurs because only one half
of the current flows down the upstream termination each time. 
The impedance of one
stripline in the BPM \vspace{-0.1in}becomes
\beq
Z_0^\parallel(\omega)=\frac12Z_c\left(\frac{\phi_0}{2\pi}\right)^2
\left(1-e^{-j2\omega\ell/c}\right)~,
\label{bpm-imp}
\vspace{-0.1in}
\eeq
where $\phi_0$ is the angle the stripline subtends at the
beam pipe axis.  The added factor, $[\phi_0/(2\pi)]^2$, indicates that only a
fraction of the image current flows across the gap into the
stripline and only this fraction sees
a gap potential.

For a separator plate, there are no terminations on either end.
As a result, while the negative image charges flow along the
beam-side of the plate, all the positive charges needed to balance
the neutrality of the plate flow along the underside of the plate.
These negative and positive charges just annihilate each other
when the beam pulse
reaches the downstream end of the plate.  
Thus there is no dissipation if the plates are considered perfectly
conducting. Therefore, the impedance in Eq.~(\ref{bpm-imp}) does not apply.
 The only contribution to the impedance come from the
resonances in the cavity gaps.  Since these resonant frequencies are high,
there is little contribution in the few hundred MHz range.

\section{IV.~~MEASUREMENT\vspace{-0.05in}}

\begin{figure}[h]
\vskip-0.08in
\centering{\epsfig{figure=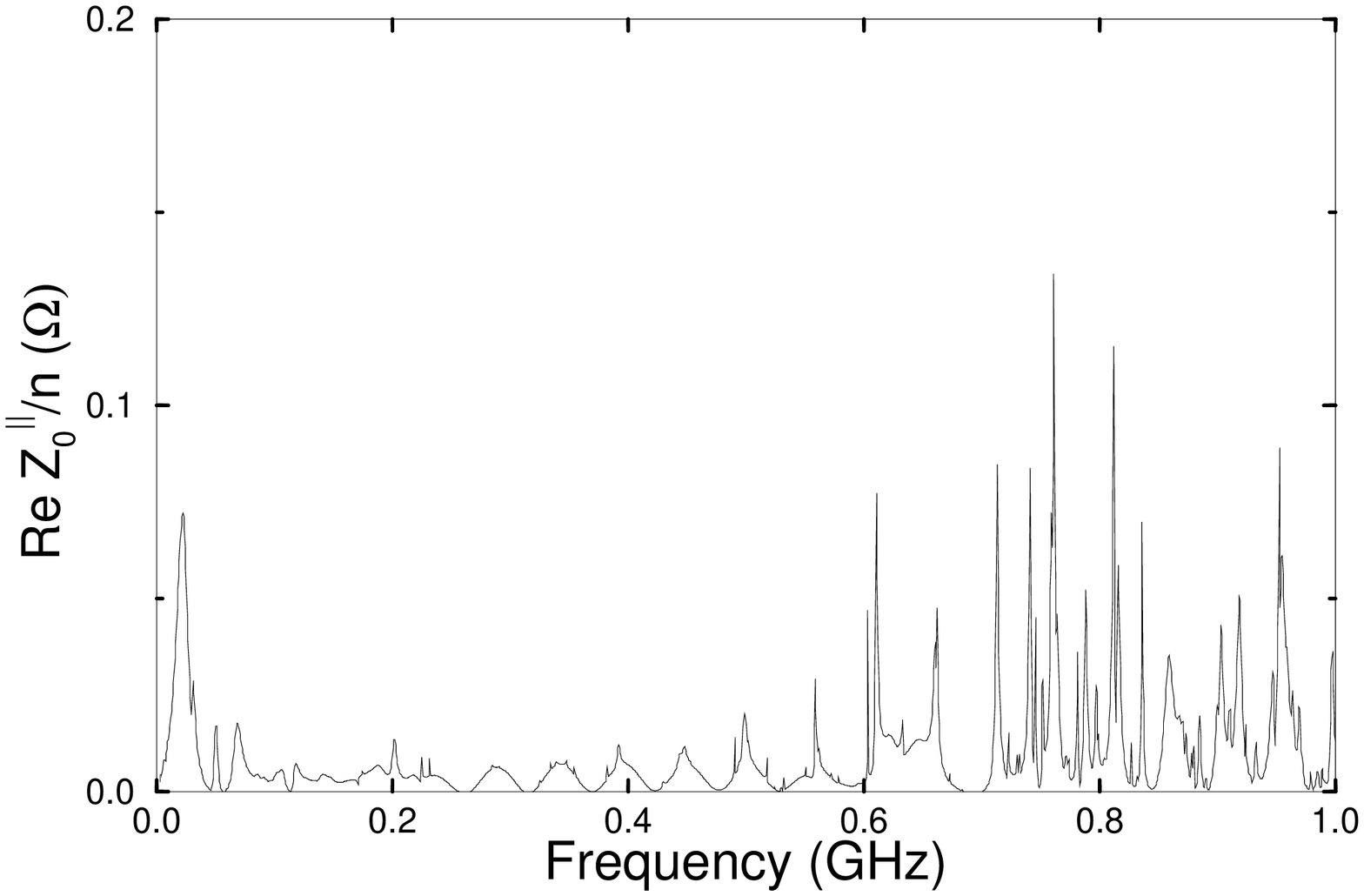,
height=3.1in,width=1.83in,angle=-90,
bbllx=180bp,bblly=35bp,bburx=565bp,bbury=647bp,clip=}}
\centering{\epsfig{figure=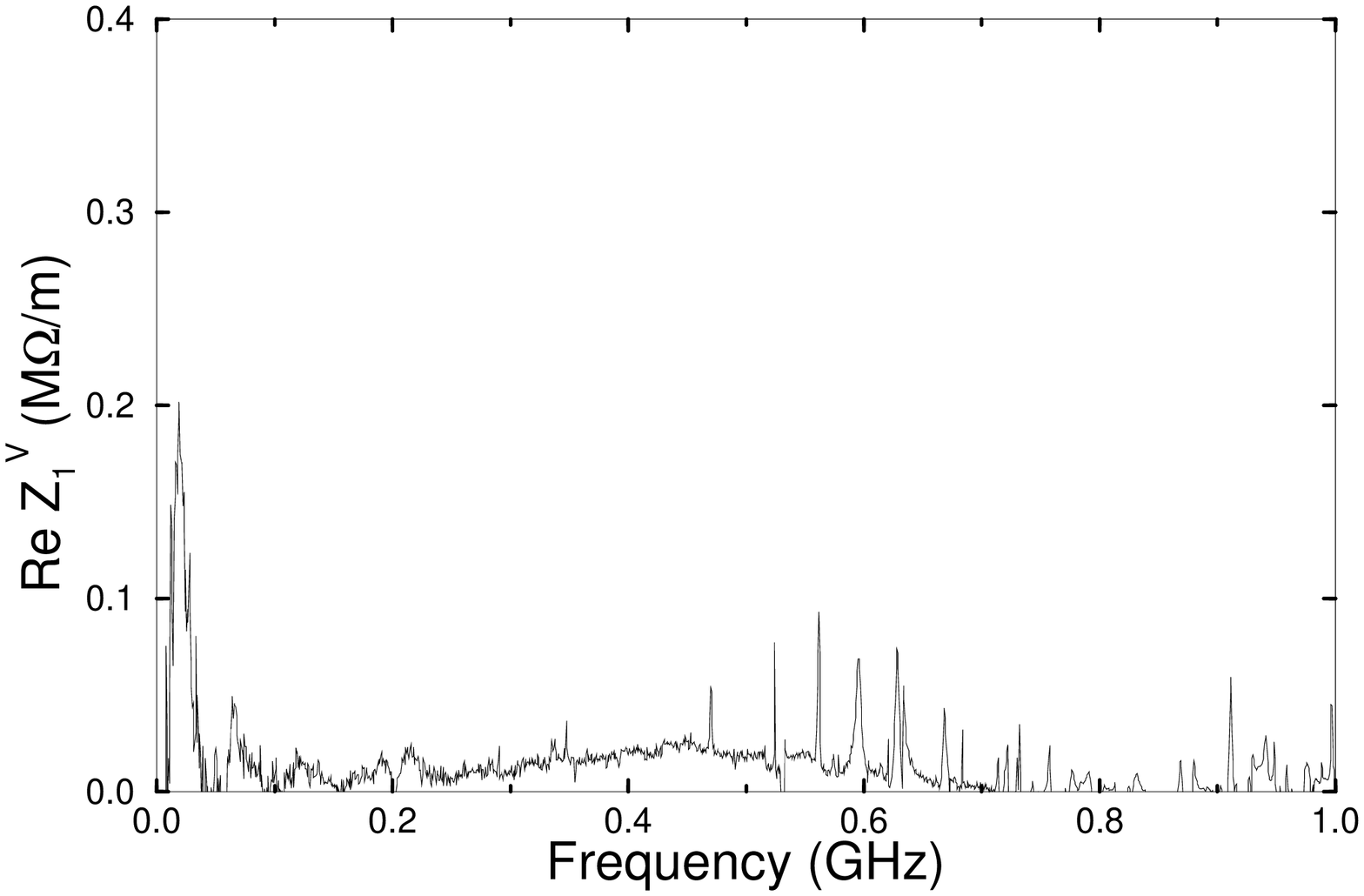,
height=3.1in,width=1.83in,angle=-90,
bbllx=180bp,bblly=35bp,bburx=565bp,bbury=647bp,clip=}}
\vskip-0.12in
\caption{\small $\Re Z_0^\parallel/n$ (top) and $\Re Z_1^V$ (bottom)
as measured by stretching wires.}
\label{trans-measured}
\vskip-0.03in
\end{figure}

The coupling impedances of a separator have recently been
measured~\cite{crisp} via the attenuation $S_{21}$
by stretching a $0.010''$ tin-plated copper wire through the separator
for the longitudinal mode and two wires for the transverse mode.
The impedances are derived 
\vspace{-0.07in}from
\beq
Z_0^\parallel=2Z_c\left(\frac1{S_{21}}-1\right)~,\quad
Z_1^V=\frac{2Z_cc\ln S_{21}}{\omega\Delta^2}~,
\vspace{-0.07in}
\eeq
where $\Delta=1$~cm is the separation of the two wires and $Z_c=50~\Omega$
is the characteristic impedance of the cables connected to the network
analyzer, to which the wires have
been matched.
In Fig.~\ref{trans-measured}, we plot\footnote{
If we plot $\Re Z_0^\parallel$ instead of $\Re Z_0^\parallel/n$,
the 22.5~MHz resonant peak will not be visible.}
the measured $\Re Z_0^\parallel/n$ and $\Re Z_1^V$.
We see that both $\Re Z_0^\parallel/n$ and $\Re Z_1^V$ contain the resonant
structures determined by MAFIA and are small in the few hundred MHz region.
However, we do see a large resonance at 22.5~MHz, 
which can be traced to
coaxial power cables attached about $12\frac12''$ from one
end of the plate through a $50~\Omega$ resistor.  Typically, the cables
are two-meter long and terminate into a 1~M$\Omega$ impedance. 
These cables extend the electric lengths of the plates and the first waveguide
mode is shifted from 54.5~MHz to 22.5~MHz.  The function of the series
$50~\Omega$ resistor is to remove any sparks if present.  Actually, this
resistor, being situated near the end of a plate, absorbs the oscillatory
current of this resonant mode.  Without this resistor, the 22.5-MHz peak
in both $Z_0^\parallel/n$ and $Z_1^V$ will be almost tripled.  
On the other hand,
these peaks will disappear if the power cables are removed.  There
are 24 separators, giving in total $\Re Z_0^\parallel/n\sim1.73~\Omega$
and $\Re Z_1^\parallel\sim 4.10$~M$\Omega$/m, which are rather large.
The longest Tevatron bunch has been $\sigma_\ell=95$~cm rms.  Considering
$\pm\sqrt{6}\sigma_\ell$, the lowest head-tail mode will have a frequency
of $c/(4\sqrt{6}\sigma_\ell)=32$~MHz.  Thus this 22.5~MHz mode may pose
a danger.  There are ways to alleviate the effect.
One is to increase the length of the power cables so as to further
reduce the resonant frequency.  A second way is to increase the damping
resistors to about $500~\Omega$,~\cite{crisp} 
hoping that the peak impedances will be
damped by a factor of ten.  The designed Tevatron rms bunch length is only
$\sigma_\ell=37$~cm.  If this shorter bunch length can be achieved,
the lowest head-tail mode will have a frequency of 82.8~MHz, too high to
be affected by the power 
\vspace{-0.1in}cables.

\section{V.~~OTHER ESTIMATION\vspace{-0.05in}}

Ref.~\cite{burov-lebedev}
suggests the vertical transverse separator \vspace{-0.1in}wake,
\beq
W_1(z)=\frac{Z_0c}{4\pi}
\frac{2}{b^2}\Big[\theta(z)-\theta(z-2\ell)\Big]\frac{\pi^2}{12}~,
\vspace{-0.08in}
\eeq
based on two plates separated by $2b=5$cm without any outer shield, 
where $Z_0$ is the free-space impedance.
The vertical transverse impedance \vspace{-0.1in}is
\beq
Z_1^V=\frac{Z_0c}{2\pi b^2}\frac{\pi^2}{12}\frac1{\omega}
\left(1-e^{-2j\omega\ell/c}\right)~,
\vspace{-0.08in}
\eeq
and becomes, at low \vspace{-0.1in}frequencies,
\beq
\Im Z_1^V=\frac{Z_0\ell}{2\pi b^2}\frac{\pi^2}{12}~,
\label{other}
\vspace{-0.08in}
\eeq
which gives the large estimate cited earlier in the Introduction.
The wake resembles the stripline BPM wake [cf Eq.~(\ref{bpm})]
with a reflected current at the downstream end of the separator plate.
As we have discussed earlier, there is no reflected current
because the positive and negative
charges created on the plate annihilate when the beam pulse crosses
the downstream separator gap.
An outer shield is very essential for a separator model, because a
waveguide/transmission line will be formed.  For the BPM, the transmission line
characteristic impedance $Z_c$ enters into
the impedance expression of Eq.~(\ref{bpm-imp}).  Without the transmission 
line, here in Eq.~(\ref{other}), the much larger
free-space impedance $Z_0$ enters 
\vspace{-0.05in}instead.

\enddocument

\section{SUBMISSION OF PAPERS}

Each author should submit all of the source files (text and figures), the
PostScript version and a hard copy of the paper.  This will allow the
editors to reconstruct the paper in case of processing difficulties and
compare the version produced for publication with the hard copy.

\section{MANUSCRIPTS}

Authors are advised to use the templates provided. Consult the individual
conference help pages if questions arise.

\subsection{General Layout}

These instructions are a typical implementation of the
requirements. Manuscripts should be prepared for one side of the paper and
have:
\begin{Itemize}
\item Either A4 (21.0 cm~$\times$~29.7~cm; 8.27~in~$\times$~11.69~in) or US
  letter size (21.6~cm~$\times$~27.9~cm; 8.5~in~$\times$~11.0~in) paper.
\item Single spaced text in two columns of 82.5 mm (3.25~in) with 5.0~mm
  (0.2~in) separation.
\item The text located within the margins specified in Table~\ref{l2ea4-t1}
  to facilitate electronic processing of the PostScript file.
\end{Itemize}

\begin{table}[hbt]
\begin{center}
\caption{Margin specifications}
\begin{tabular}{|l|c|c|}
\hline
\textbf{Margin} & \textbf{A4 Paper} & \textbf{US Letter Paper} \\ \hline
Top    & 37 mm & 19 mm (0.75 in) \\
Bottom & 19 mm & 19 mm (0.75 in) \\
Left   & 20 mm & 20 mm (0.79 in) \\
Right  & 20 mm & 26 mm (1.02 in) \\ \hline
\end{tabular}
\label{l2ea4-t1}
\end{center}
\end{table}

The layout of the text on the page is illustrated in
Fig.~\ref{l2ea4-f1}. Note that the paper's title should be the width of the
full page and that tables and figures may span the whole 170~mm page width,
if desired (see Fig.~\ref{l2ea4-f2}).

\begin{figure}[htb]
\centering
\includegraphics*[width=65mm]{JACpic_mc.eps}
\caption{Layout of papers.}
\label{l2ea4-f1}
\end{figure}

\subsection{Fonts}

In order to produce good Adobe Acrobat PDF files, the editorial staff asks
authors to use only Times (in roman, bold or italic) and Symbol from the
standard PostScript set of fonts.

\subsection{Title}

The title should use 14pt bold uppercase letters and be centered on the
page.  The names of the authors and their organizations/affiliations and
mailing addresses should be grouped by affiliation and listed in 12pt upper
and lowercase letters.

\subsection{Section Headings}

Section headings should be numbered, use 12pt bold uppercase letters and be
centered in the column. All section headings should appear directly above
the text---there should never be a column break between a heading and the
following paragraph.

\begin{figure*}[t]
\centering
\includegraphics*[width=170mm]{JACpic2.eps}
\caption{Example of full width figure showing the distribution of
problems commonly encountered  during paper processing.}
\label{l2ea4-f2}
\end{figure*}

\subsection{Subsection Headings}

Subsection headings should be numbered (e.g. \textit{2.1 General Layout}),
have 12pt italic letters with uppercase initial letters of significant
words, and be left aligned in the column. Subsection headings should appear
directly above the text---there should never be a column break between a
heading and the following paragraph.

\subsection{Paragraph Text}

Paragraphs should use 10pt roman font and be justified (touch each side) in
the column. The beginning of each paragraph should be indented
approximately 3mm (.13 in). The last line of a paragraph should not be
printed by itself at the beginning of a column nor should the first line of
a paragraph be printed by itself at the end of a column.

\subsection{Figures, Tables and Equations}

Place figures and tables as close to the place of their mention as
possible. Lettering in figures and tables should be large enough to
reproduce clearly. Use of non-approved fonts in figures often leads to
problems when the files are processed. Please use the approved fonts
whenever possible. \LaTeXe\ users---please be sure to use non bitmapped
versions of Computer Modern fonts in equations (type 1 PostScript fonts are
required, not type 3).

All figures and tables must be given sequential numbers (1, 2, 3, etc.) and
have a caption placed below the figure or above the table being described,
using 10pt roman font.

Text should not be obscured by figures.

If a displayed equation needs a number, place it flush with the right
margin of the column (see Eq.~\ref{eq:units}). Units should be written
using the roman font, not the italic font.

\begin{equation}\label{eq:units}
C_B={q^3\over 3\epsilon_0 mc}=3.54\,\hbox{$\mu$eV/T}
\end{equation}

\subsection{References}

All bibliographical and web references should be numbered and listed at the
end of the paper in a section called ``References''. When referring to a
reference in the text, place the corresponding reference number in square
brackets~\cite{exampl-ref}.

\subsection{Footnotes}

Footnotes on the title and author lines may be used for acknowledgements,
affiliations and e-mail addresses. A nonnumeric sequence of characters (*,
\dag, \ddag, \S) should be used. All other notes should be included in the
references section and use the normal numeric sequencing.

\subsection{Acronyms}

Acronyms should be defined the first time they appear.

\section{PAGE NUMBERS}

\textbf{DO NOT have any page numbers}. The editorial staff will add them
when they produce the final proceedings.

\section{TEMPLATES}

Templates and examples can be retrieved through Web browsers like Netscape
and Internet Explorer by loading to disk. See your local documentation for
details of how to do this.

Template documents for the recommended word processing software are
available from the JACoW Website~\cite{templates-ref} and exist for
\LaTeXe\ and Microsoft Word (Mac and PC) for US letter and A4 paper sizes.

Authors are required to use templates for the correct paper size.
Do not transport Microsoft Word documents across platforms, e.g.
Mac~$\leftrightarrow$~PC.

\section{CHECKLIST FOR ELECTRONIC PUBLICATION}

\begin{Itemize}
\item  Use only Times (roman, bold or italic) and Symbol fonts for
  text---10 pt minimum.
\item  Figure labelling should be in Times (roman, bold or italic) and
  Symbol fonts whenever possible---6pt minimum.
\item  Check that the PostScript file prints correctly.
\item  Check that there are no page numbers.
\item  Check that the margins are correct on the printed version. There
  may be differences of $\pm$1~mm on the margins from one printer to
  another.
\item  \LaTeXe\ users can check their margins by invoking the
  \texttt{boxit} option.
\item  Check the size of the PostScript file---an average size is about
  100--300 kbytes. If the file size is over 300 kbytes, please try to make
  it smaller (normally by reducing the complexity of the figures).
\end{Itemize}

\end{document}